\documentclass[aps,pra,twocolumn,groupedaddress,showpacs]{revtex4-1}

\usepackage{amsmath}
\usepackage{mathrsfs}
\usepackage{amssymb}
\usepackage{graphicx,epstopdf,epsfig}
\usepackage{bm}
\usepackage{color}
\usepackage{times}
\usepackage{hyperref}
\usepackage{braket}
\usepackage{multirow}
\usepackage[table]{xcolor}
\usepackage{longtable}
\usepackage{subfigure}
\usepackage{comment}

\begin{document}

\title{Measurement of Holmium Rydberg series through MOT depletion spectroscopy}

\author{J. Hostetter}
\email{hostetter@wisc.edu}
\author{J. D. Pritchard}
\author{J. E. Lawler}
\author{M. Saffman}
\email{msaffman@wisc.edu}
\affiliation{Department of Physics, University of Wisconsin-Madison, 1150 University Avenue, Madison, Wisconsin 53706}

\date{\today}

\begin{abstract}
We report measurements of the absolute excitation frequencies of $^{165}$Ho $4f^{11}6sns$ and $4f^{11}6snd$ odd-parity Rydberg series. The states are detected through depletion of a magneto-optical trap via a two-photon excitation scheme. Measurements of 162 Rydberg levels in the range $n=40-101$  yield quantum defects well described by the Rydberg-Ritz formula. We observe a strong perturbation in the $ns$ series around $n=51$ due to an unidentified interloper at 48515.47(4)~cm$^{-1}$. From the series convergence, we determine the first ionization potential $E_\mathrm{IP}=48565.939(4)$~cm$^{-1}$, which is three orders of magnitude more accurate than previous work. This work represents the first time such spectroscopy has been done in Holmium and is an important step towards using Ho atoms for collective encoding of a quantum register.
\end{abstract}


\maketitle

\section{Introduction}

Rydberg atoms have been studied extensively, yielding important information about atomic structure and ionization thresholds \cite{gallagher88,aymar96}. Rydberg atoms are attracting intense current interest in the area of quantum information processing  due to their strong dipole-dipole interactions \cite{saffman10}.  Rydberg dipole blockade provides a strong, switchable interaction between neutral atoms, allowing for the creation of  quantum gates and entanglement\cite{lukin01}.  Two-qubit gates have  so far been demonstrated in the alkali atoms rubidium \cite{isenhower10,Wilk2010} and cesium\cite{Maller2014}.  There is further interest in using the Rydberg blockade in ensembles of atoms to create a collectively encoded quantum register \cite{lukin01,brion07}.  Collective encoding is most beneficial in atoms with a large number of ground hyperfine states \cite{saffman08}.  The stable atom with the largest number of hyperfine ground states is holmium.  Its nuclear spin is $I=7/2$ and  the electronic angular momentum of the ground state is $J=15/2$, providing  128 hyperfine ground states with total angular momentum $F=4-11$. The large angular momentum arises due to the open $4f$-shell in the ground state electronic configuration. As with other rare-earth elements, this open shell structure results in an extremely complex energy spectrum that is challenging to reproduce theoretically due to strong relativistic effects and configuration
interactions \cite{biemont05}.  For Rydberg atom quantum information processing, knowledge of the Rydberg levels is important for accurate prediction of the dipole-dipole interaction strengths and sensitivity to external fields.

Studies of the Rydberg spectra of neutral rare earth elements have been limited to date, with initial measurements of the ionization potentials for the full range of lanthanides \cite{worden78} and actinides \cite{erdmann98} and energy resolved Rydberg states for La \cite{xue97}, Eu \cite{nakhate00}, Dy \cite{xu92}, Lu \cite{ogawa99}, Gd \cite{miyabe98}, Sm \cite{jayasekharan00}, Th \cite{vidolova84}, Ce \cite{vidolova97}, Yb \cite{camus80}, Ac \cite{rossnagel12}, Pu \cite{worden93} and U \cite{solarz76}. In this paper we present the first high resolution spectroscopy of the $4f^{11}6sns$ and $4f^{11}6snd$ odd-parity Rydberg states of Ho in the range $n=40-101$ using depletion measurements on a magneto optical trap (MOT). The resulting spectra are used to extract accurate values for the first ionization potential and quantum defects for the series, in addition to revealing a strong perturbation around $51s$ which is analyzed in the framework of multi-channel quantum defect theory (MQDT). Additionally, we observe a previously unpublished repumper transition from a metastable state giving significant enhancement in the MOT atom number. These measurements provide important information about the fundamental atomic structure of the open shell configuration for testing against \textit{ab initio} models.

\section{Experimental Setup}

\begin{figure}[!t]
\includegraphics[width=8.5cm]{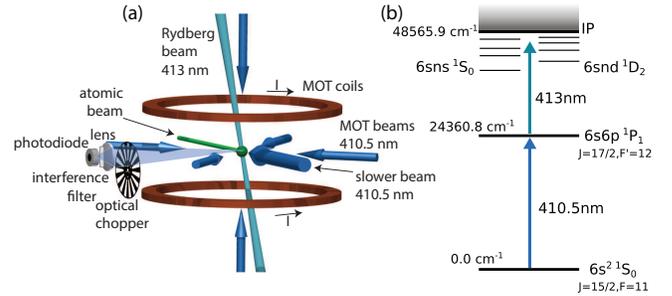}
\caption{(color online) Experimental Setup (a) A focused Rydberg excitation laser is overlapped with a Ho MOT, resulting in depletion of the equilibrium atom number on resonance. (b) Energy level diagram.  Rydberg atoms are created through two-photon excitation via $6s6p~^1P_1~F'=12$, with the first photon provided by the 410.5 nm MOT beams.\label{fig:apparatus}}
\end{figure}

Depletion spectroscopy is performed using two-photon excitation with $\sim$100 kHz linewidth lasers  of a Ho MOT via the strong $4f^{11}(4I^\circ_{15/2})6s6p(^1P^\circ_1)$ cooling transition, with a radiative line width of $\gamma/2\pi=32$~MHz. The experimental setup, shown schematically in Fig.~\ref{fig:apparatus}(a), uses the apparatus detailed in Ref.~\cite{miao14}. An atomic beam of Ho is slowed using a 400~mW counter propagating  beam, derived from a frequency doubled Ti:Sa laser, which is   detuned in the range of $-210$ to $-330~\rm MHz$ from the $4f^{11}(4I^\circ)6s^2, F=11$ to $4f^{11}(4I^\circ_{15/2})6s6p(^1P^o_1), F'=12$ transition at 410.5~nm. Atoms are loaded into a MOT formed at the intersection of three pairs of orthogonal beams with a detuning of -20~MHz and a total power of 400~mW, resulting in a saturation parameter of $I/I_\mathrm{sat}=1.06$. The equilibrium atom number is $7.5\times 10^5$ at a temperature  $<500~\mu\rm K$, with a density of $2.2\times10^{10}~{\rm cm}^{-3}$. 

  The MOT population is measured using a photodiode to monitor fluorescence. The collected fluorescence signal is amplitude modulated with an optical chopper, followed by lock-in detection, to suppress background electronic noise.   At the two-photon resonance, atoms are excited to  Rydberg states leading to increased loss of the MOT from decay into dark states and photoionization of the Rydberg atoms, reducing the equilibrium number. For each measurement, spectra are averaged over 50 repetitions and  increasing and decreasing frequency ramps of the 413 nm laser are compared to verify that resonances are observed in both scan directions as well as to account for any hysteresis in the frequency ramp.

The MOT cooling laser is stabilized to an ultra  high finesse ULE reference cavity (${\mathcal F}\sim 2\times 10^6$) mounted in vacuum and temperature controlled to $<\pm 10 ~\rm mK$, providing $<1$~MHz  frequency drift for several weeks of measurements. The short term laser linewidth after locking to the reference cavity is estimated to be $\sim 100~\rm  kHz$.  Rydberg excitation is achieved using a frequency doubled 826 nm diode laser producing 3~mW at 413 nm, which is focused to a waist of $27~\mu\rm m$ and overlapped on the MOT. The 826 nm laser is locked to a Fabry-Perot reference cavity and scanned across 1-2~GHz with a scan period of about 10 s using the cavity piezo. The Fabry-Perot cavity uses a 10 cm long  Invar spacer mounted inside a temperature controlled vacuum can. The cavity finesse is about 500 giving a cavity linewidth of about 3 MHz. The 826 nm laser is referenced to the slowly scanned cavity using a Pound-Drever-Hall locking scheme\cite{Drever1983} giving an estimated short term linewidth of 200 kHz at 413 nm. The vacuum can is temperature stabilized to better than 10 mK giving a long term frequency fluctuation of about 8 MHz at 413 nm. The combined 
frequency fluctuation of  the two lasers, which is dominated by drift of the Invar reference cavity, is thus about 10 MHz or 0.0003 cm$^{-1}$. 

The uncertainty in our determination of the energy of Rydberg levels is dominated by the uncertainty in our wavemeter measurement of the 413 nm laser light. The 410.5 nm MOT laser is referenced to an  independent measurement of the centre of gravity transition frequency from $4f^{11}6s^2, J=15/2$ to $4f^{11}6s6p, J'=17/2$ is obtained with a $1~\rm m$ Fourier Transform Spectrometer  calibrated against an Argon line using the experimental setup described in ref.~\cite{lawler04}. This gives an energy of 24360.790(1)~cm$^{-1}$, which is combined with precise measurements of the ground \cite{dankwort74,burghardt82} and excited state \cite{miao14} hyperfine splittings to yield an absolute frequency of the MOT laser  given by 730.31682(3)~THz. This value is used to calibrate the wavemeter that measure the frequency of the scanned 413 nm Rydberg laser, resulting in a total uncertainty of 200~MHz or 0.007 cm$^{-1}$ in the absolute energy of the measured levels. Details of the energy calibration procedures are provided in the Appendix.

\section{Results and Discussion}

\begin{figure}[t]
	\includegraphics{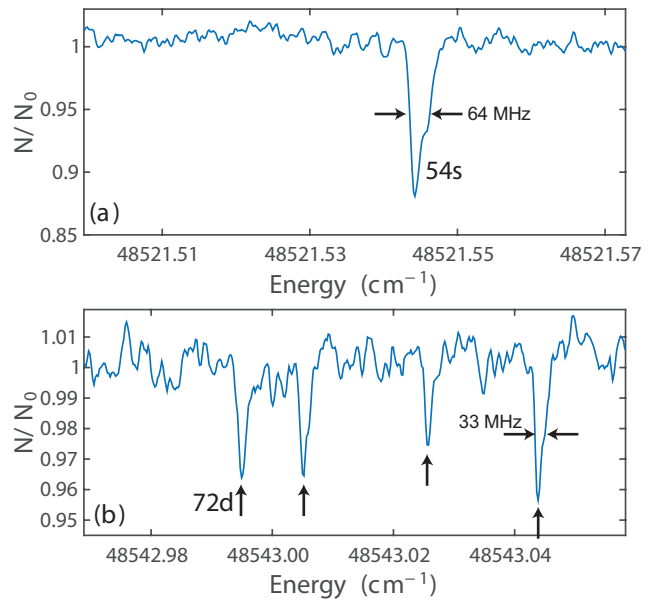}
\caption{(Color Online) MOT Depletion Spectra (a) Isolated depletion line for the $54s$ state, resolved with a 60~MHz FWHM. (b) Multiple fine structure states are resolved for the $72d$ series, with the lowest energy level used to determine the series parameters as it is typically the strongest resolved line for each $n$. \label{fig:spectra}}
\end{figure}

From the $4f^{11}6s6p, J=17/2$  state it is possible to excite the atom to the $4f^{11}6sns$ and $4f^{11}6snd$ Rydberg states, with a total of 12 series accessible. Due to parity conservation, the coupling to the triplet series is expected to be very weak, reducing this to the seven odd parity singlet states $6sns (^1S_0)$ J=15/2 and $6snd_{5/2,3/2} (^1D_2)~J=15/2,17/2,19/2$ which are observed in the experiment. Figure~\ref{fig:spectra} shows typical spectra for the $ns$ and $nd$ Rydberg series respectively, demonstrating both the relatively narrow ($\sim30-60$~MHz) spectral width of the technique as well as resolution of the fine-structure splitting of the $nd$ state resonances in (b). As the oscillator strength decreases for higher $n$ whilst the ionization rates and available dark states increase, the absolute value of the depletion is a poor indicator of absolute transition strength, however this does provide relative strengths for closely spaced fine structure transitions. The number of fine structure states resolved for the $nd$ series varies between measurements due to the finite frequency range the second photon is scanned over, but for all datasets where the full range is included between 4-6 states are resolved, limited by the signal to noise for the weaker transitions.

\begin{figure}[!t]
\includegraphics{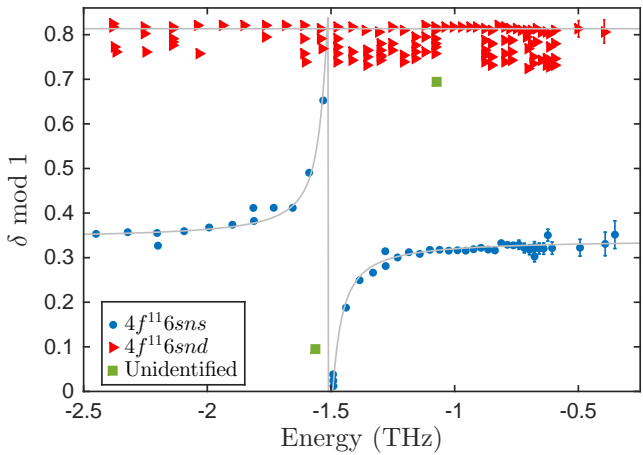}
\caption{(Color Online) Fano plot of the quantum defect modulo one versus the energy of the Rydberg states. The quantum defect is constant as is expected for the d-series (red triangles) and shows a perturbation at around n=51 in the s-series (blue circles).  Green squares indicate Rydberg levels of unknown configuration. Missing fine structure states on the $nd$ series arise due to finite measurement ranges, but have been observed for all $n$ states measured across the $\delta=0.7\ldots0.85$ range. Error bars are indicated only when larger than symbol size.\label{fig:Fano}}
\end{figure}

The energy levels $E_n$ of the Rydberg series are described by the Rydberg-Ritz formula 
\begin{equation}
E_n = E_\mathrm{IP} - \frac{\mathrm{Ry}}{[n-\delta(n)]^2},
\end{equation}
where $E_\mathrm{IP}$ is the ionization potential which represents the series limit as $n\rightarrow\infty$ and $\delta(n)$ is the quantum defect for each series. For high-lying Rydberg states the quantum defect can be assumed to be independent of $n$, allowing extraction of a precision value for $E_\mathrm{IP}$ by fitting to the series convergence \cite{worden78}. To verify which series the measured energy levels belong to, a Fano plot of $\delta$ modulo 1 against energy is used, as shown in Fig.~\ref{fig:Fano}. The measured energies then collapse into two distinct series, with the $ns$ series having a strong series perturbation around -1.5~THz and the $nd$ defects approximately constant for all $n$ for each of the different fine structure states. For the $nd$ data, the series with the largest defect is typically the strongest line and has been observed across the full range of measured energies. For this reason, only this fine structure state is used for analysis of the $nd$ defect, as the remaining satellite states are insufficiently discriminated to clearly identify their corresponding series correctly. As the strongest dipole matrix elements are for transitions to $6snd_{5/2} (^1D_2), J=19/2$, this is the most probable state being analyzed, from which we infer an inverted fine structure in the $nd$ series, as is observed in the Ho$^+$ ionic ground state. Two additional states which could not be assigned to either $ns$ or $nd$ series were observed at 48530.035(7) and 48513.837(7)~cm$^{-1}$, indicated by squares on Fig.~\ref{fig:Fano}.

The ionization potential is obtained using a least-squares fit to both the $ns$ and strong $nd$ series with $E>-1$~THz resulting in $E_\mathrm{IP}=48565.909(3)~$cm$^{-1}$, with both series converging to the $4f^{11}(^4I^\circ_{15/2})6s_{1/2},J=8$ Ho$^+$ ground state. The uncertainty in $E_\mathrm{IP}$ is an improvement of three orders of magnitude over previous measurements \cite{worden78}. In addition to obtaining the ionization potential, this fit also yields the asymptotic values of the quantum defects corresponding to 4.324(5) and 2.813(3) for the $ns$ and $nd$ series respectively. Whilst these defects are fitted modulo 1, the integer assignment is achieved by comparison to previous work \cite{fano76} which predicts that the quantum defect should be $\sim$4.3 for the $s$ series and $\sim$2.8 for the $d$ series from the variation of $\delta$ with atomic number. This is in good agreement with the defects calculated for the known energies of the $6s^2$ and $6s5d$ states.

  \begin{figure}[!t]
 \includegraphics{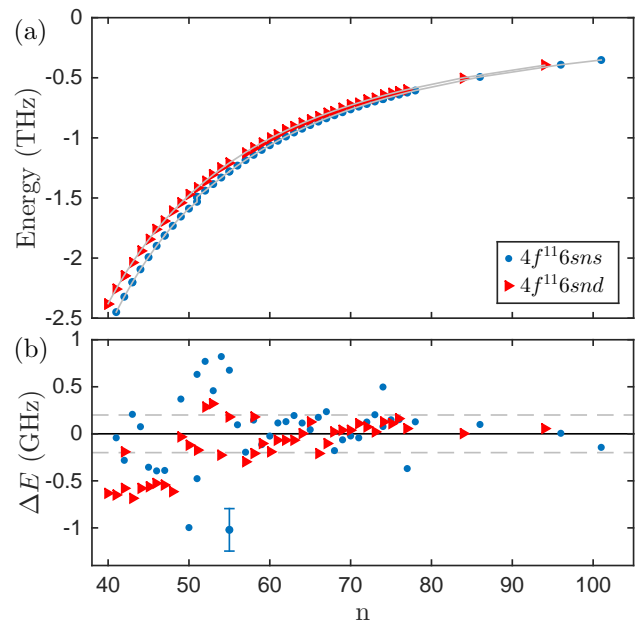}
 \caption{(Color Online) Rydberg state energy levels. (a) Measured energy levels for the $ns$ and $nd$ series, with solid lines calculated using fitted quantum defect parameters summarized in the text. (b) Fit residuals, yielding an r.m.s. residual of 200~MHz for $n>60$ away from the $51s$ series perturbation, in good agreement with the 200~MHz uncertainty (indicated by the dashed gray lines and plotted as a representative error bar).\label{fig:Residuals}}
 \end{figure}
  
For lower lying $n$ the quantum defect can be described by the Ritz formula $\delta(n)= \delta_0  + \delta_2/(n-\delta_0)^2$ \cite{ritz03} which is used to fit the $nd$ series. Unfortunately the uncertainty in $\delta_2$ is an order of magnitude larger than the best-fit value, and the series is therefore best described by a constant defect of $\delta_0=2.813(3)$. 

 Due to the strong series perturbation in the $ns$ series, the data cannot be described by a single channel quantum defect model, and instead MQDT applies \cite{seaton83}. In this framework, the perturbation can be treated as coupling between two near-resonant series, resulting in a modification of the quantum defect of \cite{seaton83,rossnagel12}
\begin{equation}
\delta(n)=\delta_0(n)-\frac{1}{\pi}\tan^{-1}\left[\frac{\Gamma/2}{(E_n-E_j)}\right],
\label{eq:mqdt}
\end{equation}
where $\Gamma$ is the spectral width and $E_j$ the energy of the interloper. Interestingly, despite the perturbation occurring below the first ionization potential associated with the Ho II ground state, the bound-bound series interaction can have a large spectral width comparable to a bound-continuum autoionization resonance \cite{aymar96}.

 \begin{table}[!t]
     \begin{ruledtabular}
     \begin{tabular}{c c c c}
$n$ & Level Energy (cm$^{-1}$) & $\delta$ & $\Delta E$ (GHz) \\\hline
41 & 48484.198 & 4.353 & -0.04 \\
42 & 48488.466 & 4.355 & -0.28 \\
43 & 48492.429 & 4.357 & 0.21 \\
*43 & 48492.537 & 4.357 & 3.46 \\
44 & 48496.074 & 4.360 & 0.08 \\
45 & 48499.442 & 4.364 & -0.36 \\
46 & 48502.578 & 4.370 & -0.39 \\
*47 & 48505.407 & 4.377 & -2.90 \\
47 & 48505.490 & 4.378 & -0.39 \\
48 & 48508.150 & 4.391 & -1.68 \\
49 & 48510.713 & 4.417 & 0.37 \\
50 & 48512.925 & 4.476 & -1.00 \\
51 & 48514.823 & 4.677 & 1.62 \\
51 & 48516.151 & 4.011 & -1.75 \\
51 & 48516.181 & 4.017 & -0.48 \\
51 & 48516.207 & 4.022 & 0.63 \\
52 & 48517.905 & 4.201 & 0.77 \\
53 & 48519.736 & 4.257 & 0.46 \\
54 & 48521.543 & 4.282 & 0.82 \\
55 & 48523.194 & 4.294 & -1.02 \\
55 & 48523.250 & 4.295 & 0.68 \\
56 & 48524.853 & 4.302 & 0.10 \\
57 & 48526.379 & 4.308 & -0.20 \\
58 & 48527.843 & 4.312 & 0.15 \\
59 & 48529.210 & 4.315 & -0.12 \\
60 & 48530.516 & 4.317 & -0.02 \\
61 & 48531.756 & 4.319 & 0.12 \\
62 & 48532.929 & 4.320 & 0.13 \\
63 & 48534.045 & 4.322 & 0.19 \\
64 & 48535.100 & 4.323 & 0.11 \\
65 & 48536.104 & 4.324 & 0.04 \\
66 & 48537.066 & 4.324 & 0.17 \\
67 & 48537.981 & 4.325 & 0.24 \\
68 & 48538.837 & 4.326 & -0.18 \\
69 & 48539.671 & 4.326 & -0.06 \\
70 & 48540.465 & 4.327 & -0.02 \\
71 & 48541.222 & 4.327 & -0.04 \\
72 & 48541.951 & 4.327 & 0.13 \\
73 & 48542.646 & 4.328 & 0.20 \\
74 & 48543.305 & 4.328 & 0.08 \\
*74 & 48543.319 & 4.328 & 0.50 \\
75 & 48543.943 & 4.328 & 0.15 \\
76 & 48544.551 & 4.329 & 0.15 \\
77 & 48545.118 & 4.329 & -0.37 \\
78 & 48545.694 & 4.329 & 0.13 \\
86 & 48549.460 & 4.331 & 0.10 \\
96 & 48552.850 & 4.331 & 0.01 \\
101 & 48554.161 & 4.332 & -0.14 \\
     \end{tabular}
     \end{ruledtabular}
\caption{Measured energies for the $4f^{11}6sns$ series, accurate to 0.007 cm$^{-1}$ with best-fit quantum defect $\delta$ determined from Eq. (\ref{eq:mqdt}) using $\delta_0=4.341(2)$, $\Gamma=6.9(3)$~GHz and $E_j=48515.47(4)$~cm$^{-1}$. The residuals relative to the Rydberg-Ritz formula are labeled $\Delta E$. Multiple $n=51$ states appear due to the series perturbation, as well as a number of additional weak doublets indicated by an asterix (*).\label{tab:ns}}
     \end{table}

The fitted defect is shown in Fig.~\ref{fig:Fano} which accurately reproduces the observed perturbations, resulting in  
parameters $\delta_0=4.341(2)$, $\Gamma=6.9(3)$~GHz and $E_j=48515.47(4)$~cm$^{-1}$. The perturbing series is unknown, but likely has the same $4f^{11}$ inner electronic configuration due to the strength of the interaction. Around this resonance the $ns$ series is observed to split into doublets, resulting in four separate states being identified as $n=51$ due to fine structure splitting of the series interloper. Weak doublets are also observed for a few other $s$ states ($n=43,47,74$), due either to  much weaker inter-series resonances or potentially singlet-triplet mixing within the $ns$ series. Above the ionization potential, we also observe a strong autoionization resonance at 48567.958(1) cm$^{-1}$ with a FWHM of 9(1) GHz. This lies 60 GHz above the first ionization potential, and thus is not an excitation to the $J = 7$ fine structure
state of the ionic core which lies  637.~cm$^{-1}$ above $E_\mathrm{IP}$ \cite{martin78}.

\begin{table*}[!t]
\begin{ruledtabular}
\begin{tabular}{c  c c c c c c c}
$n$ & Energy (cm$^{-1}$) & $\Delta E$~(GHz) & \multicolumn{5}{c}{Level Energies (cm$^{-1}$)}\\\hline
40&48486.532&-0.63& 48486.503 & 48486.720& 48486.781& &\\
41&48490.634&-0.64& 48490.700 & 48490.861& & &\\
42&48494.441&-0.19& 48494.428 & 48494.525& 48494.556& 48494.583&\\
43&48497.936&-0.68& 48498.149 & & & &\\
44&48501.200&-0.57& & & & & \\
45&48504.231&-0.55& & & & & \\
46&48507.054&-0.52& 48507.126 & & & &\\
47&48509.687&-0.55& & & & & \\
48&48512.145&-0.61& 48512.215 & 48512.295& 48512.344& &\\
49&48514.466&-0.03& 48514.563 & & & &\\
50&48516.620&-0.13& 48516.711 & 48516.740& 48516.782& &\\
51&48518.643&-0.17& 48518.752 & 48518.789& & &\\
52&48520.560&0.28& 48520.641 & 48520.642& 48520.668& 48520.702&\\
53&48522.351&0.32& 48522.398 & 48522.427& 48522.445& 48522.452& 48522.461\\
54&48524.019&-0.22& 48524.079 & 48524.107& & &\\
55&48525.622&0.17& 48525.705 & 48525.732& & &\\
56&&& 48527.144 & 48527.169& 48527.191& 48527.216&\\
57&48528.526&-0.29& 48528.563 & 48528.606& & &\\
58&48529.884&0.18& 48529.871 & 48529.906& 48529.920& 48529.930& 48529.949\\
59&48531.145&-0.10& & & & & \\
60&48532.348&-0.19& & & & & \\
61&48533.495&-0.07& & & & & \\
62&48534.581&-0.06& & & & & \\
63&48535.613&-0.07& & & & & \\
64&48536.598&-0.00& 48536.624 & 48536.643& 48536.654& 48536.669&\\
65&48537.537&0.13& 48537.562 & 48537.592& 48537.606& &\\
66&48538.417&-0.20& & & & & \\
67&48539.270&-0.10& 48539.295 & 48539.321& 48539.335& &\\
68&48540.085&0.02& 48540.107 & 48540.127& 48540.139& &\\
69&48540.860&0.03& & & & & \\
70&48541.601&0.04& & & & & \\
71&48542.310&0.10& 48542.319 & 48542.339& 48542.368& &\\
72&48542.987&0.07& 48543.005 & 48543.026& 48543.036& &\\
73&48543.634&0.02& 48543.651 & & & &\\
74&48544.259&0.13& 48544.297 & & & &\\
75&48544.854&0.11& 48544.890 & 48544.899& & &\\
76&48545.427&0.16& 48545.444 & 48545.461& 48545.469& &\\
77&48545.972&0.06& 48545.988 & 48546.007& 48546.015& &\\
84&48549.261&0.00& & & & & \\
94&48552.714&0.05& & & & & \\
     \end{tabular}
     \end{ruledtabular}
\caption{Measured states for the $4f^{11}6snd$ series, accurate to 0.007 cm$^{-1}$. The first column represents the dominant series used for determination of the quantum defect and IP for the series. The residuals relative to the Rydberg-Ritz formula are labeled 
$\Delta E$, with $\delta_0=2.813(3)$. The  remaining levels are weaker  fine structure states which could not be unambiguously assigned to a common series.\label{tab:nd}}
     \end{table*} 

Figure~\ref{fig:Residuals} shows the resulting energy residuals from the $ns$ and $nd$ fits, which with the exception of the increased error around the $51s$ series perturbation results in an r.m.s. residual of 200 MHz for $n>60$, in good agreement with the 200 MHz uncertainty in determination of the absolute energy of the levels. Absolute energies and residuals for each state are given in Tables~\ref{tab:ns} and \ref{tab:nd}, in addition to energies of the other $nd$ fine structures states.

 \begin{figure}[!t]
 \includegraphics{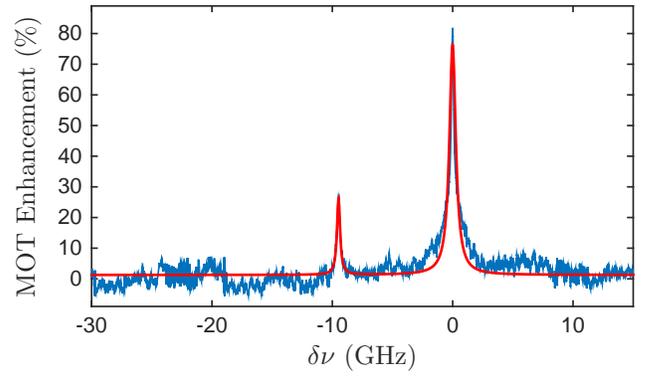}
 \caption{(Color Online) Enhancement of the MOT number as a function of  detuning from $\nu=724.2662(2)$~THz due to repumping population from a long lived metastable state back to the ground state cooling cycle, with a resolved hyperfine splitting of 9.47(1) GHz.\label{fig:ehancement}}
 \end{figure}

\section{MOT Enhancement Resonance}

Alongside the depletion resonances, a strong enhancement feature was observed resulting in up to 80\% increase in the MOT atom number at $\nu=724.2662(2)$~THz, as shown in Fig.~\ref{fig:ehancement}. The two enhancement features have a FHWM of 600 MHz and are spaced by 9.47(1) GHz, consistent with transitions between different hyperfine states. Enhancement of the MOT arises due to repumping population either from uncoupled ground state hyperfine levels or long-lived metastable excited states back into the $F=11$ ground state which is cooled in the MOT. The peak enhancement wavelength is independent of the MOT detuning, verifying that this effect arises from a  single photon repump transition which can be exploited to create large atomic samples. The closest matching transition from published line data is from $4f^{10}(^5I_8)5d_{5/2}6s^2, J=17/2$ to $4f^{10}5d6s6p, J=17/2$ \cite{kroger97}, which has a frequency of 724.350~THz, within 10~GHz of the measured resonance.

\section{Summary}
In summary we have measured 162 odd-parity Rydberg states belonging to the $4f^{11}6sns$ and $4f^{11}6snd$ series using MOT depletion spectroscopy, providing the first energy resolved Rydberg spectra for neutral Ho. Analysis of the measured levels yields a significantly improved determination of the first ionization potential of $E_\mathrm{IP}=48565.939(4)$~cm$^{-1}$, as well as  asymptotic quantum defects for the $ns$ and $nd$ series of 4.341(2) and 2.813(3) respectively. These data provide an important reference for testing \textit{ab initio} theories predicting the energy levels of complex atoms. The observation of regular $ns$ and $nd$ series without strong perturbations for most $n$ in the range $40\le n \le 101$ suggests that we can expect to find long lived Rydberg states suitable 
for creating collectively encoded quantum registers \cite{saffman08}.
Determination of Rydberg lifetimes and interaction strengths will be the subject of future work.  In addition to Rydberg levels, a strong repump transition has been identified enabling a significant increase in MOT atom number. This will be useful for preparing large, dense atomic samples as a starting point for creation of a dipolar Bose-Einstein condensate \cite{lahaye09,lu11,aikawa12} exploiting the large 9~$\mu_\mathrm{B}$ magnetic moment.

\begin{acknowledgments}
This work was supported by funding from NSF grant 1404357.
\end{acknowledgments}

\pagebreak

\newpage

\appendix

\section{Energy Level Calibration}

 \begin{figure}[!t]
\includegraphics{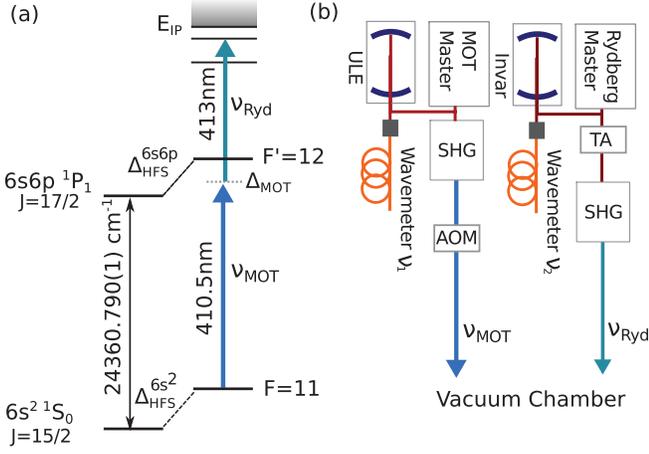}
\caption{Holmium Energy Levels. (a) MOT and Rydberg laser transitions with respect to the ground state energy level and hyperfine splitting. (b) Laser setup showing cavity locks, wavelength measurement and additional frequency shift from AOMs on the cooling laser. A wavemeter is used to measure the laser frequencies $\nu_{1,2}$ before they are doubled and sent to the atoms.\label{fig:cal}}
\end{figure}

To provide accurate energy levels of the measured Rydberg states we use an independent measurement to determine the absolute frequency of the MOT laser, which is stabilized to the TEM$_{00}$ mode of a high finesse ULE cavity providing a stable long term frequency reference. Figure~\ref{fig:cal}(a) shows the relevant energy levels and splittings used in the experiment, whilst the laser setup is shown in (b). The centre of mass frequency for the $4f^{11}6s^2~^1S_0~J=15/2$ to $4f^{11}6s6p~^1P_1~J=17/2$ transition is determined by the 1m Fourier Transform Spectrometer at the National Solar Observatory using the setup detailed in ref.~\cite{lawler04}. High resolution spectra from a Ho-Ar hollow cathode lamp are recorded, using lines in the well known ArII series \cite{learner88,whaling02} for calibration to give a transition frequency of $E_{6s^2\rightarrow6s6p}=24360.790(1)~\rm cm^{-1}$ which is 0.02 $\rm cm^{-1}$ less than the value in the NIST tables \cite{martin78}.

For the cooling transition from $F=11$ to $F'=12$, the hyperfine splitting of the ground and excited states are calculated from measurements of the hyperfine constants giving  $\Delta_\mathrm{F=11}^{6s^2} = 20.589319(1)$~GHz \cite{dankwort74,burghardt82} $\Delta_\mathrm{F'=12}^{6s6p} = 19.33(1)$~GHz \cite{miao14}. The frequency of the MOT transition can then be calculated from
\begin{equation}
\nu_\mathrm{MOT}=E_{6s^2\rightarrow6s6p} + \Delta_\mathrm{F'=12}^{6s6p} - \Delta_\mathrm{F=11}^{6s^2}+\Delta_\mathrm{MOT},
\end{equation}
where $\Delta_\mathrm{MOT}=-22(2)$~MHz is the detuning from resonance determined from spectroscopy of the atomic beam, giving $\nu_\mathrm{MOT}=730.31682(3)$~THz. Accounting for the double pass acousto-optic modulator (AOM) at 50~MHz and frequency doubling of the SHG, the frequency of the master Ti:Sa laser locked to the cavity is given by $\nu_\mathrm{ref}=365.15836(2)$~THz. For each measurement, the wavelength of this laser ($\nu_1$) is recorded on a wavemeter to determine the wavemeter offset $\delta\nu=\nu_\mathrm{ref}-\nu_1$. The Rydberg energy levels are then calculated from measuring the Rydberg master laser frequency ($\nu_2$) before the doubling cavity on the same wave meter, resulting in $\nu_\mathrm{Ryd}=2(\nu_2+\delta\nu)$, with the absolute energy above the ground state given by
\begin{equation}
E_\mathrm{Ryd} = \Delta_\mathrm{F=11}^{6s^2}+\nu_\mathrm{MOT}+\nu_\mathrm{Ryd},
\end{equation}
resulting in a total uncertainty of 200~MHz on the final energy reading.

\end{document}